# Electromagnetic potentials and Aharonov-Bohm effect

Alexander Ershkovich

Department of Geophysics and Planetary Sciences, Tel Aviv University, Ramat Aviv, 69978, Israel, alexer@post.tau.ac.il

**Abstract**

Hamilton-Jacobi equation which governs classical mechanics and electrodynamics explicitly depends on the electromagnetic potentials (**A**, φ), similar to Schroedinger equation. We derived the Aharonov-Bohm effect from Hamilton-Jacobi equation thereby having proved that this effect is of classical origin. These facts enable us to arrive at the following conclusions: a) the very idea of special role of potentials (**A**, φ) in quantum mechanics (different from their role in classical physics) lost the ground, and becomes dubious, as this idea is based on the Aharonov-Bohm effect, b) failure to find any signs of a special role of these potentials in the appropriate experiments (*Feinberg*, 1963) is thereby explained, and c) discovery of classical analogues of the Aharonov-Bohm effect (*Berry et al.*, 1980) is also explained by a classical nature of this effect. Elucidation of the "unlocal" interaction problem was made.

It is generally believed that the magnetic potential **A** which in classical physics is considered as an auxiliary, mathematical concept, in quantum mechanics, owing to Aharonov-Bohm effect [1], acquires a physical meaning. Thus, this effect resulted in an idea of special role of electromagnetic potentials in quantum mechanics (different from their role in classical physics). Explicit dependence of Schroedinger equation on these potentials (rather than on corresponding fields), according to [1], corroborates this idea, which became popular (for more details see [2, 3] and references therein).

We drew attention [4] to the fact that Hamilton-Jacobi equation which governs the classical mechanics and electrodynamics depends explicitly on the potential **A** (similar to Schroedinger equation), thereby pointing out that the Aharonov-Bohm effect is of classical origin. We also have shown [5] that this effect follows from classical Hamilton-Jacobi equation for the action $S$

$$\frac{\partial S}{\partial t} + H = 0 \qquad (1)$$

where

$$H = \frac{1}{2m}(\nabla S - q\mathbf{A})^2 + q\varphi \qquad (2)$$

is the Hamilton function. Here $m$ is the mass of a particle with the charge $q$ in the electromagnetic field with potentials (**A**, φ). The characteristics of the equation (1), of course, are the Hamilton equations of motion. But equation (1) possesses another peculiarity: it describes a wave-particle duality in classical physics, thereby being, in fact, a source of quantum mechanics: in the quasi-classical approximation Schroedinger equation results in Hamilton-Jacobi equation (1). Equation (1) coincides with the eikonal equation for the wave phase $\psi$ in the approximation of geometrical optics [6]:

$$\frac{\partial \psi}{\partial t} + \omega = 0 \qquad (3)$$

Both approximations above hold with small wavelengths $\lambda$, namely, with $kL\gg1$, where $k = 2\pi/\lambda$, $L$ is the characteristic scale of the relevant nonuniformity. Experiments which revealed the Aharonov-Bohm effect well satisfy the condition $kL\gg1$.

Of course, roots of the wave particle duality (described by equation (1)) are located in the close affinity between Fermat and Hamilton principles which explains the optics-mechanical analogy.

Thus, the solutions of equations (1) and (3) are proportional: $S = Const\ \psi$. With $H = \hbar\omega$ one obtains $S = \hbar\psi$. The eikonal $\psi = S/\hbar$ may be considered as the phase of the de Broglie wave, with $\omega = H/\hbar$ and $\mathbf{k} = m\mathbf{v}/\hbar$. In case of the charge $q$ in the magnetic field the action $S$ is known to acquire an additional term

$$S_{in} = q \int \mathbf{A} d\mathbf{r} \qquad (4)$$

which describes an interaction of the charge $q$ with the field. Thus the phase shift $\Delta\psi$ accumulated due to charge-field interaction equals

$$\Delta\psi = S_{in}/\hbar = \frac{q}{\hbar} \int \mathbf{A} d\mathbf{r} \qquad (5)$$

We arrive at the Aharonov-Bohm effect [1]. Schroedinger equation has not been used. The classical roots of this effect are obvious. These roots explain the existence of classical analogues of the Aharonov-Bohm effect. Berry et al. [7, 8] found one of such classical effects on the water surface by using hydrodynamics-electrodynamics analogy. Of course, the effect [1] is a quantum one (as it depends on $\hbar$) but not an effect of quantum mechanics: the relation $H = \hbar\omega$ used above was known long before the quantum mechanics creation by Heisenberg and Schroedinger in 1926.

The assertion that Aharonov-Bohm effect is indicative of special role of the potential $\mathbf{A}$ underwent a careful and comprehensive examination in [9]. Feinberg [9] arrived at the conclusion that the experiments suggested by Aharonov and Bohm [1] do not contain anything that has not been known from the classical electrodynamics. Several examples given by Feinberg show that some real, observed physical phenomena in classical physics depend explicitly on the potentials ($\mathbf{A}$, $\varphi$) rather than on the corresponding fields. It is quite obvious why Feinberg did not find any signs of special role of these potentials in the Aharonov-Bohm effect: the effect [1] is of classical origin, it is derived from the equation (1). But if so, this effect cannot be a ground for an idea of special role of electromagnetic potentials in quantum mechanics. Are there any other reasons to support this idea? It seems dubious.

Schroedinger equation for the charged particle in the electromagnetic field is (e.g. [10])

$$i\hbar \frac{\partial \Psi}{\partial t} = \hat{H}\Psi \qquad (6)$$

where the Hamilton operator is

$$\hat{H} = \frac{1}{2m}(\hat{P} - q\mathbf{A})^2 + q\varphi \qquad (7)$$

$\Psi$ is the wave function, the momentum operator $\hat{P} = -i\hbar\nabla$. Rewrite the expression (7) as

$$\hat{H} = \frac{1}{2m}(\hat{P}^2 - 2q\mathbf{A}\hat{P} + i\hbar q\nabla\mathbf{A} + q^2\mathbf{A}^2) + q\varphi \qquad (8)$$

where the relation $\hat{P}\mathbf{A} - \mathbf{A}\hat{P} = -i\hbar\nabla\mathbf{A}$ (associated with the Heisenberg uncertainty prinsiple) has been employed. It is obvious that the terms with $\nabla\mathbf{A}$ and $\mathbf{A}^2$ in the equation (8) have nothing in common with the idea on special role of potentials in quantum mechanics. In addition, the Coulomb gauge, $\nabla\mathbf{A}=0$ is required for commutativity of the momentum operator $\hat{P}$ with the potential $\mathbf{A}$, in order to obey the Heisenberg uncertainty principle (see e.g. [10] for details).

Comparing classical Hamilton function (2) with equation (8) we arrive at the conclusion that the potentials ($\mathbf{A}$, $\varphi$) participate in quantum mechanics in the same way as in the classical electrodynamics, namely, as vector and scalar functions of coordinates, respectively, rather than as differential operators. Therefore, one cannot expect that in quantum mechanics they play any special role, different from that in classical physics.

In conclusion, we would like to determine the physical meaning of the potential $\mathbf{A}$ as the momentum of charge-field interaction. Indeed, a particle with the charge $q$ and the momentum $m\mathbf{v}$ in the magnetic field with the potential $\mathbf{A}$ acquires an additional momentum, $q\mathbf{A}$, so that $\mathbf{P} = m\mathbf{v}+q\mathbf{A}$. Thus, $\mathbf{A}$ is the momentum of interaction of the test particle carrying the unit charge, with the field. Naturally, the same physical meaning is given by equation (4):

$$\mathbf{A} = \frac{1}{q}\left(\frac{\partial S}{\partial \mathbf{r}}\right)_{in} = \frac{1}{q}\mathbf{P}_{in} \qquad (9)$$

where $\mathbf{P}_{in} = \mathbf{P} - m\mathbf{v}$. The physical meaning of the potentials ($\mathbf{A}$, $\varphi$) remains the same in quantum mechanics (compare equations (2) and (7)).

**Updated** (April 10, 2013)

There is an another problem associated with Aharonov-Bohm effect. This is so called electron-solenoid unlocal interaction. Indeed, the magnetic field $\mathbf{B} \neq 0$ only within a long solenoid. Outside $\mathbf{B} = 0$ but $\mathbf{A} \neq 0$. Nevertheless, Aharonov-Bohm effect is indicative of electron-solenoid interaction there [1]. Vivid discussion on "unlocal" interaction (through the potential $\mathbf{A}$) is lasting already more than half of century (e.g. [2], [8], [11]). Feinberg [9] was the first to show that interaction between the electron current $\mathbf{j} = q\mathbf{v}$ and solenoid due to Faraday induction has to be taken into account. As a result, the field $\mathbf{B} \neq 0$ arises outside the solenoid, so that the Lorentz force $q\mathbf{v}\times\mathbf{B} \neq 0$. Boyer [12] independently arrived at this conclusion by using the momentum conservation in Maxwell electrodynamics. Naturally, an action term $S_{in} = q\int \mathbf{A}d\mathbf{r}$ (equation (4)) includes the Faraday induction effect. That is why Aharonov-Bohm effect (5) has been obtained from classical Hamilton-Jacobi equation (1). Notice that a steady-state description $\partial/\partial t = 0$ happened to be insufficient in order to solve the problem above. Indeed, the Biot-Savart law (its differential form is $\nabla \times \mathbf{B} = \mathbf{j}$), generally speaking, is inconsistent with the third Newton's law. But according to Noether theorem, conservation laws describe most fundamental characteristics of our space-time world, and, as such, these laws must be fulfilled in a correct theory.

Thus, as shown by Feinberg [9] and Boyer [12], a "mystery" above is resolved in the framework of classical electrodynamics. Then, according to Occam's razor there is no need to introduce a mystical "unlocal" interaction.

**Acknowledgement.** Author is grateful to S. A. Trigger who attracted his attention to the paper [9] by E.L. Feinberg, as well as for useful discussions**.**